# Highly Conducting Spaced TiO$_2$ Nanotubes Enable Defined Conformal Coating with Nanocrystalline Nb$_2$O$_5$ and High Performance Supercapacitor Applications

Selda Ozkan,[1,‡] Nhat Truong Nguyen,[1,‡] Imgon Hwang,[1] Anca Mazare,[1] Patrik Schmuki[1,2,*]

S. Ozkan, N. T. Nguyen, I. Hwang, Dr. A. Mazare, Prof. Dr. P. Schmuki

[1]Department of Materials Science, Institute for Surface Science and Corrosion WW4-LKO, Friedrich-Alexander University, Martensstraße 7, D-91058 Erlangen, Germany

[2]Chemistry Department, Faculty of Sciences, King Abdulaziz University, 80203 Jeddah, Saudi Arabia Kingdom

[‡]These authors contributed equally to this work.

*Corresponding author. E-mail: schmuki@ww.uni-erlangen.de

In this work, we report on the electrochemical behavior of nitrided spaced TiO$_2$ nanotubes conformally coated with a nanocrystalline Nb$_2$O$_5$ layer and find for these hierarchical structures an excellent supercapacitor performance. Highly aligned conductive 1D electrodes were obtained by a three step process: *i)* growth of self-organized nanotubes with defined and adjustable intertube spacing, *ii)* conformal Nb$_2$O$_5$ decoration in the tube interspace (while providing full electrolyte access to the entire active area), and *iii)* high temperature nitridation. Key is the growth of a nanotube array with regular tube-to-tube interspacing that enables an optimized decoration with secondary materials such as Nb$_2$O$_5$. We observe an increase in electrode capacitance from 158 µF cm$^{-2}$ for bare TiO$_2$ NTs, to 1536 µF cm$^{-2}$ for TiO$_2$/Nb$_2$O$_5$ NTs, to finally 37 mF cm$^{-2}$ for Nb$_2$O$_5$ decorated and then nitrided nanotubes. This drastic increase can be ascribed firstly to the defined spacing established between the tube arrays that then allows for a conformal coating with a secondary active coating. Secondly, nitridation causes a drastic increase of the electron conductivity of the entire scaffold and thus reduces resistive losses.

Due to increasing energy demands, energy conversion and storage technologies have been intensively researched and developed in recent years. Supercapacitors are considered a very promising candidate for electrochemical energy storage due to their high power density, long cycle life and fast discharge rate.[1–3] According to the charge storage mechanism, supercapacitors are generally classified into two categories: electric double layer capacitors (EDLCs) and pseudocapacitors (PCs).[4,5] EDLCs are based on non-faradaic charge separation at the electrode/electrolyte interface while pseudocapacitors are dominated by faradaic process on the electrode. Pseudocapacitors have increasingly attracted interest, since they allow intrinsically for a higher specific capacitance and energy density, provided that fast and reversible redox reactions can be established at the electrode.[4]

For such pseudocapacitors, to maintain an overall charge neutrality, ion-intercalation into the material takes place as redox switching occurs. Ion intercalation and the electrode conductivity (as it allows fast and efficient electron transfer), as well as a large surface area and non-diffusion limiting geometries are key aspects that define the effectiveness of a supercapacitive system. From the materials side, $RuO_2$ represents the gold standard for supercapacitor applications due to a fast and reversible intercalation combined with a high electron conductivity.[4,6,7] However, the high cost of $RuO_2$ restricts its commercial use. Therefore, a wide range of metal oxide materials such as $MnO_2$, $TiO_2$, $SnO_2$, $Nb_2O_5$, $V_2O_5$, $MoO_3$, $NiO$, or carbides–nitrides have been considered as potential electrode materials.[8–15] To meet the criteria of a large surface area, electron transport and diffusion geometries, various nanoscale structures such as nanosponges, mesosponges, fibers, nanotubes, nanowires, nanobelts, nanoplatelets, nanoribbons, nanorods, nanoparticles and nanochannels were intensively explored.[7,13,14,16–19] Specifically, $TiO_2$ nanotubes obtained by electrochemical anodization offer a high specific surface area combined with directionality for ion and charge transport,[8,16,17,20–22] as these oxide structures are vertically aligned, directly on the Ti metal back contact.

In general, self-ordered $TiO_2$ nanotube arrays grow close-packed, i.e. have no or only very narrow intertube spacing. This limits the use of such $TiO_2$ NTs as a scaffold for building up defined hierarchical structures, e.g. by the deposition of nanoparticles. Furthermore, $TiO_2$ as a scaffold is far from an ideal capacitor due to the low electronic conductivity of $TiO_2$. In the present work, we address these points by establishing self-organizing conditions that lead to a defined tube spacing and by converting the $TiO_2$ structure to a highly conducting scaffold by nitridation.

As a decoration material, we selected $Nb_2O_5$ since it can provide a high switching capability with different valence states and it shows a high chemical and electrochemical inertness.[11] When $Nb_2O_5$ is converted to nitrides, the Nb-N system largely consists of nitrogen-rich phases, e.g. $Nb_4N_5$ and $Nb_5N_6$ – containing several high valence (+5) Nb ions, that are promising candidates for electrochemical capacitors.[23] Moreover, Nb-N compounds are reported to provide a high cyclic stability.[24,25]

In order to build this hierarchical structure we grow spaced (SP) TiO$_2$ NTs by electrochemical anodization of a titanium foil in a di-methyl sulfoxide (DMSO) electrolyte and then coat the nanotubes conformally with Nb$_2$O$_5$. To overcome the intrinsic low electrical conductivity of TiO$_2$, we use a high temperature NH$_3$ treatment that converts the oxides to highly conducting nitrides. Such TiO$_2$ nanotube layers show a drastic increase of the capacitance, in comparison to bare (non-decorated or reference) or non-nitrided layers.

**Figure 1** shows schematically the assembly of the TiO$_2$/Nb$_2$O$_5$/NH$_3$ electrode used in this work: (*i*) self-organizing anodization to spaced tubes, (*ii*) Nb$_2$O$_5$ decoration, and lastly (*iii*) nitridation under NH$_3$ flow. After a set of exploring experiments (see also SI) we selected spaced NTs that are grown in a DMSO electrolyte consisting of HF, NH$_4$F and H$_2$O using anodization at 30 V for 4 h (Figure 1b and S1). The resulting as-formed NTs are amorphous, have a diameter of 119 ± 14 nm, a length of ~10 µm and a bottom tube-to-tube spacing of 91 ± 29 nm (more details regarding the formation of spaced nanotubes are given in the SI and in Figure S2). Prior to the Nb$_2$O$_5$ decoration, the nanotubes were annealed to anatase (in air, 450 °C). Nb$_2$O$_5$ decoration is then accomplished by spin coating with niobium (V) chloride. After drying, the electrodes were sintered (in air, at 500 °C), in order to crystallize the Nb$_2$O$_5$ coating.

To assess the optimal amount of Nb$_2$O$_5$ loading, both the classic close-packed (CP NTs) and spaced nanotubes (SP NTs) were decorated with different amounts of Nb$_2$O$_5$ (see Figure S3a-b). The amount of Nb$_2$O$_5$ can be controlled by repeating the spin-coating procedure, e.g. 1 time (1t), 4 times (4t) and 10 times (10t). For close-packed NTs, starting from 1t, the Nb$_2$O$_5$ layer partially clogs the tube openings (see Figure S3a), whereas for spaced NTs, even for 10t loading, there is no clogging and the Nb$_2$O$_5$ covers the outer wall of tubes very uniformly from top to bottom, as well as inside the tubes, close to bottom (Figure S3b). TEM images (Figure 1c) confirm that the Nb$_2$O$_5$ layer is deposited to both the outer and inner side of the spaced NTs. The elemental analysis of close-packed and spaced NTs with different $Nb_2O_5$ loadings is shown in Table S1. After the $NH_3$ treatment, the oxides are partially converted to nitrides and the resulting decoration consists of nanocrystalline Nb-N and Nb-O mixed phases, as shown in Figure 1d, S4 and S5, for a more detailed discussions see the SI. Although spaced NTs have overall less surface area than the close-packed NTs, the spacing in between NTs enables an easy access for the decoration with metal oxide from top to the bottom, as well as inside the nanotubes. This provides for an overall increase in the surface area that can be decorated while still maintaining access for electrolyte solutions.

The dependence of the electrochemical performance on the $Nb_2O_5$ loading for both close-packed and spaced NTs is shown in Figure S6a-b. Electrochemical characterization was performed in 1 M $H_2SO_4$ using a three-electrode set-up – with a Pt counter electrode and Ag/AgCl reference electrode. While the capacitance of the 10 µm close-packed NTs is $\approx$ 583 µF cm$^{-2}$, spaced NTs have an areal capacitance of 158 µF cm$^{-2}$. The capacitance of SP NTs further increases with 1t $Nb_2O_5$ decoration up to 1382 µF cm$^{-2}$; and with 4t (or 10t) up to 1536 µF cm$^{-2}$ (or 1880 µF cm$^{-2}$) (Figure S6b).

To convert the $TiO_2$ and $Nb_2O_5$ to the more conductive titanium nitride-niobium nitride semi-metallic phases, the scaffolds were annealed in $NH_3$ at temperatures between 750 °C to 950 °C. In order to optimize the loading and the nitridation treatment, we evaluated the structures by cyclic voltammograms (CV) for bare (non-decorated) and $Nb_2O_5$ decorated nanotubes after using different nitridation temperatures. Firstly, for a fixed nitridation temperature (900 °C), different loadings were tested (see Figure S7). From these results 4t loading showed an optimized current density, while at a higher loading amount (10t) the current density was found to be drastically reduced.

For extracting an optimal nitridation procedure, we evaluated the CV curves (100 mV s$^{-1}$) of the bare and 4t $Nb_2O_5$ decorated nanotubes, nitrided at different temperatures that is from 750 to 950 °C (Figure S8). For bare nitrided nanotubes ($TiO_2$/$NH_3$), 750 and 850 °C provide a better performance, whereas for $Nb_2O_5$ decorated nanotubes ($TiO_2$/$Nb_2O_5$/$NH_3$), 850 and 900 °C result in the highest values of the current density. Nevertheless, comparing the influence of the decoration for each nitridation temperature, for the 750 °C treatment the performance is reduced with increasing the $Nb_2O_5$ decoration, for 850 °C it slightly increased, while for 900 °C, there is a noticeable increase. If the duration of the 850 °C $NH_3$ treatment is further increased, e.g. to 15 min or 1 h, the performance drastically decreases as illustrated in Figure S9a-b. This is due to the fact that an extension of the $NH_3$ treatment duration or an increase in temperature (e.g. to 950 °C), leads to a decay of the nanotube structure (deep cracks in the layer, visible by eye). Additionally, the layer loosens its adhesion to the Ti substrate and as a result the electrical properties worsen. In this respect, the optimal temperature is 900 °C as it leads to a better performance while maintaining the tubular structure of the nanotubes – see SEM cross-section images of the spaced nanotubes for reference $TiO_2$ (Figure S1a) and for $TiO_2/Nb_2O_5/NH_3$ scaffold (Figure 1e and 1e1-3).

XRD analysis shows the presence of the niobium (V) oxide before the $NH_3$ treatment in Figure 2a. Annealed $TiO_2$ NTs crystallize in the form of anatase (JCPDS 21-1272) and rutile (JCPDS 21-1276). Before nitridation, $Nb_2O_5$ has a pseudo-hexagonal crystal structure (indicated with TT-$Nb_2O_5$, JCPDS 28-0317) with diffraction peaks at 22.58°, 28.60°, 36.74° and 46.16°. For $TiO_2$ NTs annealed in $NH_3$, a partial conversion to TiN or $Ti_2N$ is obvious, this depending on the annealing temperature, i.e. only TiN (JCPDS 87-0631) phase at 750 °C, TiN-$Ti_2N$ (JCPDS 17-0386 for $Ti_2N$) at 850, 900 and 950 °C (Figure 2a, S10). Additionally, the ammonia treatment reduces the TT-$Nb_2O_5$ crystal structure partially to a NbO cubic phase (JCPDS 77-0015). This is in line with literature that reports high temperature treatments under reductive conditions to convert the insulator $Nb_2O_5$ to the conductive NbO or $NbO_2$ phases.[25] It is noteworthy that niobium oxide can be converted to NbN (JCPDS 043-1420), evident from XPS analysis (Figure 2b-e), while its diffraction peak overlaps with the Ti peak at ~38.5° (JCPDS 044-1294). The polycrystalline nature (mixture of oxide-nitride) of the niobium decoration is confirmed by high-resolution TEM images, and the lattice fringes for the NbN-NbO phases are shown in Figure 1d. The distance between the lattice fringes results as 0.355 nm ($d_{(440)}$ = 0.349 nm, JCPDS 043-1420) for NbN, 0.243 nm ($d_{(111)}$ =

0.243 nm, JCPDS 077-0015) for NbO and 0.364 nm for $Nb_4N_5$ ($d_{(101)}$ = 0.364 nm, JCPDS 074-0606), see Figure 1d and S5.

For the $TiO_2/Nb_2O_5/NH_3$ electrode treated at 900 °C (in $NH_3$), the relative amount of the niobium, titanium and nitrogen was assessed by EDX measurements (Figure S11a). The reference NTs have a Ti:O atomic ratio of 0.57 (36.6:63.4), but with further treatments ($TiO_2/Nb_2O_5/NH_3$) very low Ti:O ratios are obtained (38.7:4.7), with 27.5 at% N and 29.1 at% Nb. Thus, the low amount of oxide corroborates the conversion of titanium and niobium (V) oxides ($TiO_2/Nb_2O_5$) to suboxides and corresponding nitrides, as evident from XRD.

Moreover, the chemical composition of the $TiO_2$ nanotubes at various stages was evaluated by X-ray photoelectron spectroscopy (XPS). Figure 2b-e and S11b-c show results for the bare (reference, ref) $TiO_2$, $Nb_2O_5$ decorated ($TiO_2/Nb_2O_5$, 4t) and the 900 °C ammonia treated ($TiO_2/Nb_2O_5/NH_3$) nanotubes. The NTs decoration with $Nb_2O_5$ is confirmed by the Nb3d doublet peaks corresponding to the niobium (V) oxide, $Nb3d_{5/3}$ at 207.3 eV, $Nb3d_{3/2}$ at 210.0 eV – see Figure 2b and S11c. The incorporation of nitrogen into the oxides as a result of the ammonia treatment is evident in the N1s peak at ≈ 397 eV (Figure 2c). Please note that the nitrogen content in the bare and $Nb_2O_5$ decorated nanotubes is very small and is mostly due to N pick-up from the environment. After the ammonia treatment, additional peaks are observed in the Nb3d spectra (Figure 2b) and are attributed to the reduction of the oxide to niobium oxy-nitrides (NbON, peaks at 206.2 and 208.9 eV, respectively) or suboxides[26] and to the formation of niobium nitride ($Nb_xN_y$, peaks at 204.7 and 207.4 eV, respectively)[26] – an example of peak fitting is given in Figure 2d. The calculations reveal that from the 6.56 at% Nb, 1.85 at% is in the form of NbN and 0.95 at% is present as NbON (the rest of 3.76 at% is as niobium oxide). The formation of titanium nitride and oxy-nitrides is also observed in the Ti2p peaks (Figure S11b, 2e). The temperature of the ammonia treatment influences the NbN and NbON content, as evident from the 850 °C treatment, where 1.25 at% NbN and 0.99 at% NbON were detected.

As the ammonia treatments for 10 min at 850 °C and 900 °C show the highest current densities (CVs in Fig. 3a-b), we investigated different scan rates (from 1 mV s$^{-1}$ to 200 mV s$^{-1}$) for samples treated at these temperatures, as shown in Figure 3 and S12. These findings demonstrate that TiO$_2$/Nb$_2$O$_5$/NH$_3$ electrodes treated at 850 or 900 °C show not only a better performance than the bare nanotubes (see Figure S10c) but also that for the decorated nanotubes, with an increase in higher rate, the maximum current density becomes higher for the 900 °C treated sample. Figure 3c shows the areal capacitance versus scan rate of the TiO$_2$/Nb$_2$O$_5$/NH$_3$ electrode treated at 900 °C for 10 min and an areal capacitance of 37 mF cm$^{-2}$ is measured at a scan rate of 1 mV s$^{-1}$. The nanotubular electrode with the Nb$_2$O$_5$ decoration (TiO$_2$/Nb$_2$O$_5$/NH$_3$) shows a higher capacitance and a better rate capability compared to the bare nanotubes (TiO$_2$/NH$_3$), as illustrated in Figure 3d (see Figure S13 for galvanostatic charge-discharge results). Therefore, the niobium-oxy-nitride nanocrystalline decoration improves not only the capacitance but also the rate capability of the electrode. The niobium-oxy-nitride decoration builds a connected network on the conductive Ti$_x$N$_y$ framework for the electrochemical reactions, that overall improves the capacitance. Furthermore, the Nb-N system improves the stability of the electrode, as shown in Figure 3d.

To evaluate the electrochemical behavior of the electrodes prepared at 900 °C, electrochemical impedance spectroscopy (EIS) measurements were performed at 0.3 V with an amplitude of 10 mV. Figure 3e shows the Nyquist plots of the reference TiO$_2$, TiO$_2$/NH$_3$ and TiO$_2$/Nb$_2$O$_5$/NH$_3$ with an optimized decoration. For all samples, there is a small semicircle at high frequencies that is an indication of low charge transfer resistance (R$_{ct}$) and thus of a high conductivity. The slope increase (at low frequencies) reflects the ion diffusion in the electrolyte to the electrode interface. At very low frequencies, an ideal capacitor electrode usually exhibits an almost vertical line on the Nyquist plot and a phase angle close to 90° on the Bode plot, whereas for pseudocapacitors it can vary.[23,27–29] In the present work, the slope of the Nyquist plots is close to 90° for the reference TiO$_2$ and it deviates from 90° for TiO$_2$/Nb$_2$O$_5$/NH$_3$ electrode. Accordingly, in the Bode plots, phase angles for the reference TiO$_2$ is at around 87-88°, that is close to an ideal capacitor, while for TiO$_2$/Nb$_2$O$_5$/NH$_3$ electrodes it is ≈ 80° (see Figure S14). As the observed phase angle of the TiO$_2$/NH$_3$ and TiO$_2$/Nb$_2$O$_5$/NH$_3$ electrodes deviates from 90° at low frequency, this additionally supports a pseudocapacitive nature. Clearly, the pseudocapacitance response comes mainly from the NH$_3$ treatment, as the reference (i.e. anatase) TiO$_2$ nanotubes show nearly an ideal capacitive behavior.

Additionally, two-point conductivity measurements demonstrate the drastic effect of nitridation on the conductivity (Figure 3f). The drastically increased values can be ascribed to the conversion of TiO$_2$ and Nb$_2$O$_5$ to semi-metallic phases. Please note the fact that for both temperatures of the ammonia treatment, i.e. for 850 °C and 900 °C, Nb exists in the form of NbO, NbN and Nb$_4$N$_5$ (as previously discussed). The conversion of the Ti-O or Nb-O phases to nitrides becomes more dominant with increasing temperatures and accordingly the electrical properties further improve; reported electrical resistivities of TiN and NbN are 27 μΩ cm and 60 μΩ cm respectively.[22] The formation of TiN and NbN is beneficial for the electrical behavior, however, it is well known that TiN is not stable in aqueous or under electrochemical oxidizing conditions.[30] NbN particularly is reported as a stable electrode in aqueous electrolyte solutions.[24] Thus, TiN electrodes can be improved (in terms of stability) by combining TiN with more stable carbon materials,[31] metal oxides, or polymers, or using NbN composite nitrides.

Overall, the present work thus shows that the nitridation of a hierarchical interspaced nanotube array that is conformally decorated by Nb$_2$O$_5$, under optimized conditions, can achieve remarkable supercapacitor performance. While the bare anatase NTs behave like an ideal double layer capacitor with a 158 μF cm$^{-2}$ capacitance, the functionalized electrodes (TiO$_2$/Nb$_2$O$_5$/NH$_3$), decorated with an optimized loading and after an ammonia treatment at 900 °C, have an areal capacitance of 37 mF cm$^{-2}$ and behave as a dual pseudo capacitance and as an EDLC. The dual nature mainly arises from the TiN, Ti$_2$N Nb$_4$N$_5$ and NbN semi-metallic phases. Furthermore, the present study demonstrates the possibility to decorate the inner and outer walls of spaced nanotube arrays (exploiting particularly the interspace), to alter and significantly enhance the functional features of TiO$_2$ nanotube arrays. We believe that the concepts shown in this work can be extended to a wide range of functional applications where localized harvesting or junction combinations with TiO$_2$ NTs are beneficial.

**Experimental Section**

Prior to anodization, 0.1 mm thick Ti foils (99.6 % pure tempered annealed, ADVENT) were degreased by sonication in acetone, ethanol, and distilled water, respectively, and dried in nitrogen stream. To fabricate spaced nanotubes, anodization was performed in dimethyl sulfoxide (DMSO) electrolyte with additions of 3 wt% HF (40 % Sigma Aldrich), 1 wt% $H_2O$, and 0.3 wt% $NH_4F$ at 30 V (room temperature) for 4 h. The classic close-packed NTs were formed in 1 M $H_2O$ and 0.1 M $NH_4F$ containing ethylene glycol (EG) electrolyte at 50 V for 1 h. Anodization was carried out in a two-electrode configuration with Pt as cathode and Ti substrate as anode (IMP-Series Jaissle Potentiostat). After anodization, the samples were immersed in ethanol overnight and dried in $N_2$ stream. Before $Nb_2O_5$ decoration, the nanotubular layers were annealed at 450 °C (air) for 1 h using a Rapid Thermal Annealer (RTA). Niobium (V) chloride (99 % Sigma Aldrich) is used as a precursor to prepare 0.1 M suspensions in absolute EtOH (99 %, VWR Chemicals). $Nb_2O_5$ was decorated using a BLE Delta 10 spin coater at 2000 rpm for 30 s (repeated from 1 time to 10 times). Subsequently, nanotubes with $Nb_2O_5$ decoration ($TiO_2/Nb_2O_5$) were annealed at 500 °C for 30 min, in a tube furnace (Heraeus RO 7/50) in air. $TiO_2/Nb_2O_5$ samples were further treated in $NH_3$ flux at temperatures between 750 °C to 950 °C for 10 min ($TiO_2/Nb_2O_5/NH_3$).

The morphology was characterized using a field-emission scanning electron microscopy (FE-SEM S4800 Hitachi) coupled with an energy-dispersive X-ray detector (EDX, Genesis 4000).

X-ray diffraction (XRD, X'pert PhilipsMPD with a Panalytical X'celerator detector, Germany) was performed using graphite monochromized CuKα radiation (Wavelength1.54056 Å). Chemical composition was analyzed by X-ray photoelectron spectroscopy (XPS, PHI 5600, spectrometer, USA) using AlKα monochromatized radiation. Further morphological and structural characterizations were carried out with a CM 30 TEM/STEM (Philips).

The electrochemical tests were conducted in 1 M $H_2SO_4$ using Autolab/PGSTAT30. The cyclic voltammetry (CV) measurements were performed with a scan rate between 1 mV $s^{-1}$ to 200 mV s$^{-1}$. The galvanostatic charge-discharge measurements were performed at 0.25 mA cm$^{-2}$, 0.64 mA cm$^{-2}$, 0.96 mA cm$^{-2}$, 1.3 mA cm$^{-2}$ in the voltage range of 0 to 0.6 V. The impedance (EIS) measurements were conducted at 0.3 V in the frequency range from 10 mHz to 100 kHz at amplitude of 10 mV.

**Supporting Information** Supporting Information is available from the Wiley Online Library or from the author.

**Acknowledgements**
The authors acknowledge the ERC, the DFG, the DFG "Engineering of Advanced Materials" cluster of excellence and DFG "funCOS" for financial support. Shiva Mohajernia is acknowledged for her help with the experiments.

Received: ((will be filled in by the editorial staff))
Revised: ((will be filled in by the editorial staff))
Published online: ((will be filled in by the editorial staff))

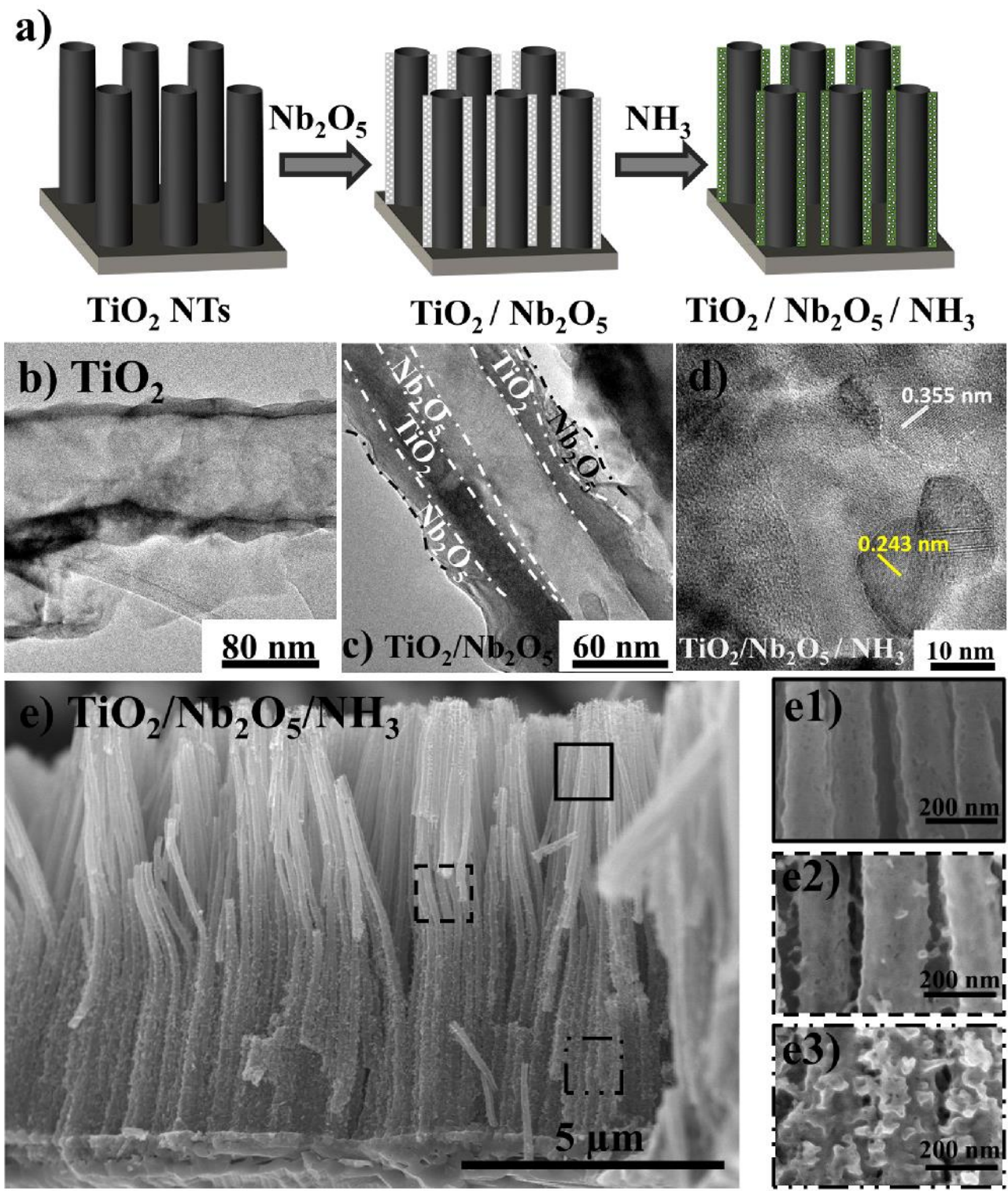

**Figure 1. a)** Schematic drawing of reference $TiO_2$, $TiO_2/Nb_2O_5$, $TiO_2/Nb_2O_5/NH_3$ NTs. TEM image of **b)** reference (bare, ref) $TiO_2$, **c)** $TiO_2/Nb_2O_5$ NTs. **d)** TEM image of the Nb-O-N nanocrystalline structure at high magnification. **e)** SEM side view image, **(e1-3)** high magnification cross-section images from top to bottom of $TiO_2/Nb_2O_5/NH_3$ (4t) treated at 900 °C.

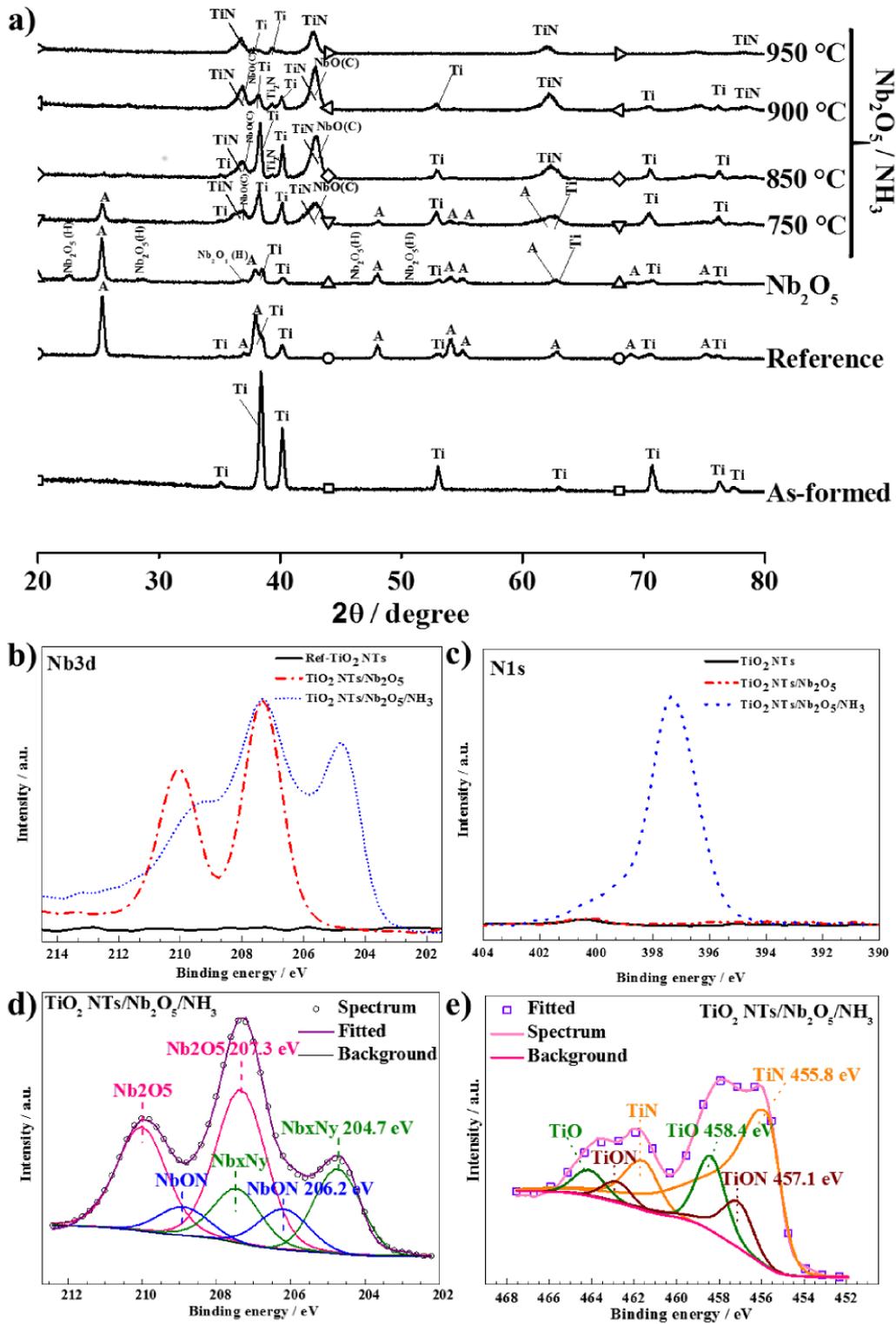

**Figure 2. a)** X-ray diffraction patterns of as-formed $TiO_2$, reference (bare, ref) $TiO_2$, $TiO_2/Nb_2O_5$ (4t) and $TiO_2/Nb_2O_5/NH_3$ (annealed in $NH_3$ from 750 °C to 950 °C for 10 min). XPS analysis for ref $TiO_2$, $TiO_2/Nb_2O_5$ and $TiO_2/Nb_2O_5/NH_3$ nanotubes under optimized conditions (900 °C for 10 min) shows **b)** fitted Nb3d peaks and, **c)** fitted N1s peaks. Peak fitting of the $TiO_2/Nb_2O_5/NH_3$ electrode prepared under optimized conditions (900 °C for 10 min) **d)** Nb3d peaks and **e)** Ti2p peaks.

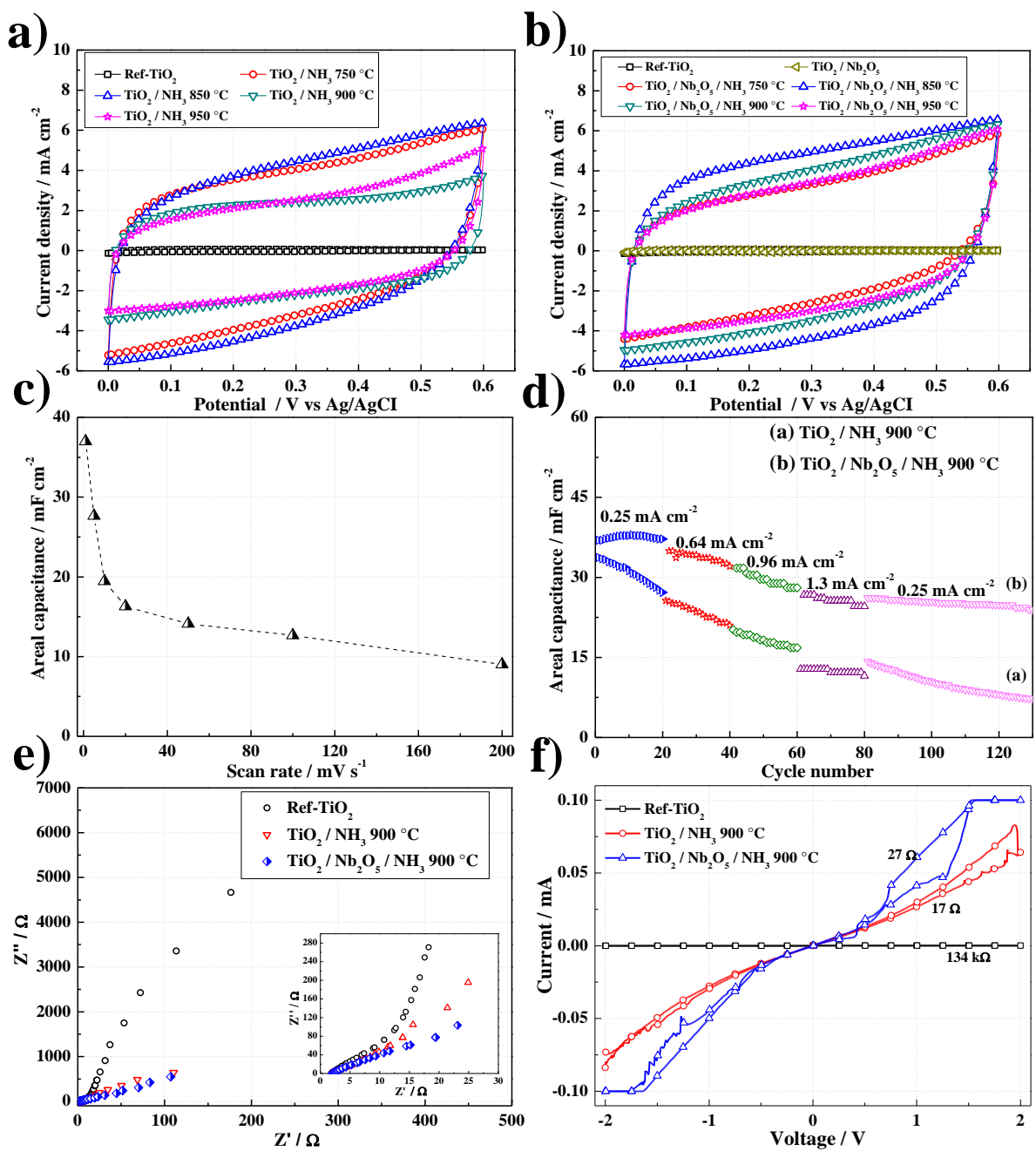

**Figure 3.** Cyclic voltammetry (CV) curves of NTs treated from 750 to 950 °C at a scan rate of 100 mV s$^{-1}$ **a)** TiO$_2$/NH$_3$, **b)** TiO$_2$/Nb$_2$O$_5$/NH$_3$ NTs. For optimized conditions (NH$_3$ treated at 900 °C for 10 min): **c)** Areal capacitance versus scan rate plot for TiO$_2$/Nb$_2$O$_5$/NH$_3$ (calculated from CV curve recorded at a scan rate of 1 mV s$^{-1}$). **d)** Rate capability of TiO$_2$/NH$_3$, TiO$_2$/Nb$_2$O$_5$/NH$_3$ (calculated from the discharge curve of galvanostatic-charge-discharge profile). **e)** Nyquist plot for ref TiO$_2$, TiO$_2$/NH$_3$, TiO$_2$/Nb$_2$O$_5$/NH$_3$ at 0.3 V in 1 M H$_2$SO$_4$. **f)** 2-point conductivity measurement of ref TiO$_2$, TiO$_2$/NH$_3$, TiO$_2$/Nb$_2$O$_5$/NH$_3$ NTs.

**Table of Contents**

**Establishing self-organized spacing between TiO$_2$ nanotubes** allows for highly conformal Nb$_2$O$_5$ deposition that can be adjusted to optimized supercapacitive behavior.

**Keyword:** anodization, spaced nanotubes, TiO$_2$, Nb$_2$O$_5$, supercapacitors

Selda Ozkan, Nhat Truong Nguyen, Imgon Hwang, Anca Mazare, Patrik Schmuki[*]

**Highly Conducting Spaced TiO$_2$ Nanotubes Enable Defined Conformal Coating with Nanocrystalline Nb$_2$O$_5$ and High Performance Supercapacitor Applications**

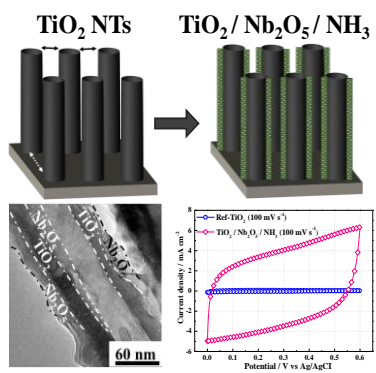